\documentclass[preprint,12pt]{elsarticle}

\usepackage{amssymb}
\usepackage{graphicx}
\usepackage{color}
\usepackage{colordvi}

\begin{document}


\begin{frontmatter}

\title{Nuclear fragmentation reactions in extended media studied with Geant4 toolkit}

\author[FIAS,INR]{Igor~Pshenichnov\corref{cor1}}
\author[FIAS,INR]{Alexander~Botvina}
\author[FIAS,KI]{Igor~Mishustin}
\author[FIAS]{Walter~Greiner}

   \address[FIAS]{Frankfurt Institute for Advanced Studies, J.-W. Goethe 
University, 
                60438 Frankfurt am Main, Germany }
   \address[INR]{ Institute for Nuclear Research, Russian Academy of Science, 
117312 Moscow, Russia}
   \address[KI]{Kurchatov Institute, Russian Research Center, 123182 Moscow, 
Russia}

\cortext[cor1]{Corresponding author: pshenich@fias.uni-frankfurt.de}

\date{\today}

\begin{abstract}
It is well known from numerous experiments that nuclear multifragmentation is a dominating 
mechanism for production of intermediate-mass fragments in nucleus-nucleus collisions 
at energies above 100 $A$ MeV.  
In this paper we investigate the validity and performance of the Fermi break-up model and 
the statistical multifragmentation model implemented as parts of the Geant4 toolkit. 
We study the impact of violent nuclear disintegration reactions on the depth-dose 
profiles and yields of secondary fragments for beams of light and medium-weight nuclei 
propagating in extended media. Implications for ion-beam cancer therapy and 
shielding from cosmic radiation are discussed.
\end{abstract}

\end{frontmatter}

\section{Introduction}

The passage of energetic nuclei through various materials is a subject 
of experimental and theoretical investigations for several decades. 
In particular, the energy deposition by light nuclei in living tissues is of a primary 
importance for ion-beam cancer therapy~\cite{Scholz:2000,Amaldi:Kraft:2005}.  
Several Monte Carlo particle transport codes like 
SHIELD-HIT~\cite{Gudowska:et:al:2004}, 
PHITS~\cite{Sato:Sihver:Iwase:2005,Zeitlin:Sihver:et:al:2008,Sato:Kase:Watanabe:2009}, 
FLUKA~\cite{Sommerer:etal:2006,Sommerer:etal:2009}, 
MCNPX~\cite{Titt:etal:2008,Stankovskiy:etal:2009} 
and Geant4~\cite{Paganetti:etal:2004,Jiang:Paganetti:2004,Kase:etal:2006,Sarrut:Guigues:2008}
are used to simulate the interaction
of protons and nuclei with homogeneous phantoms made of tissue-like materials,
voxelized phantoms based on 3D CT images of patient's body,  as well as with
beam-line elements, e.g., nozzles and collimators.

As demonstrated in Refs.~\cite{Pshenichnov:etal:2005,Pshenichnov:etal:2006,Pshenichnov:etal:2007,MCHIT}
by calculations with the Monte Carlo Model for Heavy-Ion Therapy (MCHIT),
the recent versions of the Geant4 toolkit~\cite{Agostinelli:etal:2003,Allison:etal:2006}
can be successfully used for simulations of carbon-ion therapy. 
However, until now the projectile nuclei heavier than carbon 
have received  much less attention in studies with Geant4.
Recently we have performed a comparative study of the depth-dose distributions 
for beams of various light and medium-weight nuclei in water using the Geant4 
toolkit~\cite{Pshenichnov:etal:2008}.

A proper description  of physical and biological processes induced by light and heavy nuclei 
in extended media is important also for present and future exploration of space.
This includes the design of shielding elements  
to be used during future interplanetary missions~\cite{Benton:Benton:2001,Townsend:2005}, 
reliable estimations of the doses acquired by astronauts in long space flights,
evaluations of radiation conditions at Moon and other space 
colonies. As shown in Ref.~\cite{Durante:Cucinotta:2008}, the Galactic Cosmic Rays (GCR) present 
one of the main dangers for long-term human activities in space. The GCR present 
a very broad spectrum of nuclei, from hydrogen to iron, with 
kinetic energies from 10 MeV to 100 GeV per nucleon.
The energy distribution has a maximum around 1 GeV per nucleon, however, 
the tails can reach 1 TeV per nucleon. It was demonstrated~\cite{Durante:Cucinotta:2008} 
that despite of their small fraction in GCR ($\sim 10^{-3}$), the biological impact of 
heavy nuclei like Fe on living tissues is very strong, comparable with the impact of protons.
Therefore, it is important to extend the capabilities of 
particle transport codes to the domain of heavy projectile nuclei, as it was done recently, 
for example, for the PHITS~\cite{Sato:Sihver:Iwase:2005,Zeitlin:Sihver:et:al:2008} and 
MCNPX~\cite{James:etal:ANS:2006,James-Moyers:etal:2006} codes.
 
During propagation of high-energy nuclei in extended media a certain 
fraction of them undergoes inelastic interactions with target nuclei.
It is quite common to assume that this process consists of three stages:
(1) the initial fast stage of reaction when nuclei 
interact strongly with each other; (2) preequilibrium stage when
fast particles leave a highly excited nuclear system; and (3) de-excitation of
the equilibrated nuclear residues by evaporation of nucleons and light clusters or by
breaks up into several fragments.

Most dynamical models used to describe the initial stage of
nucleus-nucleus collisions come to the conclusion that after a time interval of a
few tens of fm/c fast particles escape, and the evolution
of the remaining nuclear system changes its character. Because of
intensive interaction between nucleons the residual nuclei evolve toward
statistical equilibrium. If the excitation energy is low, the hot compound nucleus
undergoes de-excitation by evaporating nucleons and light clusters.
At higher excitation energies the hot residual nucleus 
expands and breaks up into hot primary fragments, which later on undergo de-excitation by
evaporating nucleons and light clusters. It is interesting to note that 
a wide variety of secondary nuclei, including exotic nuclei, e.g. hypernuclei, can be
created in the Earth's atmosphere by energetic cosmic nuclei. 
However, it is very difficult to detect these products of nuclear reactions.

In our studies with Geant4~\cite{Pshenichnov:etal:2005,Pshenichnov:etal:2006,Pshenichnov:etal:2007} 
the initial stage was simulated by means of the Light-Ion Binary 
cascade model (G4BinaryLightIonReaction)~\cite{Geant4:Physics:Manual:BIC}.
This model treats a nucleus-nucleus collision as a sequence of individual nucleon-nucleon
collisions in the region where the colliding nuclei overlap with each other.
The production of secondary hadrons, e.g. pions, in nucleon-nucleon collisions 
is also taken into account. Since this model is a straightforward extension of the hadron-nucleus Binary
cascade model~\cite{Wellisch}, it does not include collective phenomena such as
compression of nuclear matter or collective flow of secondary products. 
Due to these simplifications the model developers suggested that it is 
relevant to nucleus-nucleus collisions where at least one of the colliding nuclei 
is not much heavier than carbon nucleus.  In general, the Monte Carlo simulations for 
ion-therapy with beams of light 
nuclei~\cite{Pshenichnov:etal:2005,Pshenichnov:etal:2006,Pshenichnov:etal:2007} 
propagating in tissue-like materials are within the scope of the Binary cascade model, as 
mostly H, C, and O nuclei are involved. 
The Light-Ion Binary cascade model is recommended to be applied for projectile nuclei with energies 
between 50~$A$ MeV and 10~$A$ GeV~\cite{Geant4:Physics:Manual:BIC}. The primary  
excitation energy of a nuclear residue left after the cascade stage of reaction is defined 
by the numbers of excitons,i.e. particles (captured nucleons) and holes 
(nucleons removed from the nuclear core).

The Wilson abrasion model (G4WilsonAbrasionModel) is available as a part of Geant4 
toolkit~\cite{QinetiQ:Manual,Geant4:Physics:Manual:Abr} as an alternative approach for
simulating inelastic nucleus-nucleus collisions. It is dealing with 
calculating volumes of nuclear residues (spectators) and their excitation energies
rather than with detailed consideration of nucleon-nucleon collisions in the overlap (fireball) zone.
For the sake of simplicity the angular distribution of the nucleons abraded from a nucleus is 
assumed to be isotropic in the reference frame of this nucleus. The momentum distributions of such 
nucleons are simulated according to a phenomenological expression with parameters depending on the value
of Fermi momentum of the corresponding nucleus~\cite{Geant4:Physics:Manual:Abr}. 
The model has no restrictions on the masses of colliding nuclei, but the production of secondary 
hadrons, e.g. mesons, in the overlap zone is ignored. Therefore, the predictions 
of the Wilson abrasion model become less and less accurate
with increasing beam energy above the pion production threshold. 
This reduces the energy domain of applicability of 
the Wilson abrasion model as compared to the Light-Ion  Binary cascade model. 

In the abrasion model the excitation energy of nuclear
residues comes from two sources: (1) from excess of their surface energy associated 
with their non-spherical shape and (2) from the energy deposited by abraded nucleons 
passing through them. As well established, several de-excitation mechanisms are responsible for
de-excitation of nuclear residues: emission of nucleons and light clusters, 
nuclear fission and multifragmentation. 
As follows from the statistical model analysis~\cite{Bondorf:etal:1995},  
evaporation dominates at low excitation 
energies, less than 3 MeV per nucleon. It is characterized by modest changes in the charge 
and mass of the initial residual nucleus, as only few neutrons, protons and alphas are emitted in a 
relatively long time scale. Nuclear fission is a dominant de-excitation mechanism only for
very heavy nuclear residues like Th or U. 
The multifragment break-up of nuclei or multifragmentation becomes
important at excitation energies above 3 MeV per nucleon and is characterized by a relatively 
short time scale of about 100 fm/c~\cite{Bondorf:etal:1995}. As established by many experiments, 
this is a dominating process for production of intermediate mass fragments. 
In recent years it has been demonstrated by numerous calculations that this process can be well
described within the statistical approaches such as Fermi break-up model for light nuclei
and Statistical Multifragmentation Model (SMM) for medium and heavy nuclei~\cite{Bondorf:etal:1995}.
In Ref.~\cite{Botvina:Mishustin:2006}, the contribution of multifragmentation to the total 
reaction cross section was estimated at the level of 10-15\% for hadron-nucleus collisions, 
and about twice as large for nucleus-nucleus collisions. 

To the best of our knowledge, only the SHIELD-HIT code and the Geant4 toolkit take into
account multifragmentation of medium-weight and heavy nuclei in transport 
calculations for extended media. Due to radical changes in mass and charge of the projectile 
nucleus following multifragmentation, the ionization energy loss changes dramatically. Therefore, 
inclusion of nuclear multifragmentation in calculations of ion energy deposition in extended media
will lead to changes in depth-dose distributions 
as well as in yields of secondary fragments produced in nuclear fragmentation reactions. 
We believe that the role of violent disintegration of highly excited nuclei should be carefully 
evaluated for a set of practical applications which deal with energetic ion transport in extended media,  
especially for cancer therapy and cosmic radiation protection. 

The paper is organized as follows. First, we give a brief
description of the Fermi break-up model for decay of light nuclei 
(G4FermiBreakUp) in Sec.~\ref{FermiDescription}, and, second,  
the statistical multifragmentation model (G4StatMF) in Sec.~\ref{SMMDescription}. 
Both models are parts of the Geant4 toolkit. 
Then we report the results of stand-alone tests of these models. 
Validation checks of G4FermiBreakUp and 
G4StatMF are made in  Sec.~\ref{FermiValidation} and Sec.~\ref{SMMValidation}, respectively.
We simulate multifragment break-up of several hot nuclear systems with specified 
excitation energies. The results are compared with predictions of the FORTRAN-77 
version of the SMM~\cite{Bondorf:etal:1995}, which also
includes the Fermi break-up model.
In Sec.~\ref{YieldsC12} we calculate the depth-dose distributions for carbon ions in water, as well as 
the depth-yield distributions of secondary fragments produced by such nuclei. 
The calculational results are compared with corresponding experimental data. 
The depth-dose distributions for iron nuclei in water are also 
calculated and compared with experimental data in Sec.\ref{IronInWater}.
Finally, in the same section we demonstrate the impact of multifragmentation reactions on 
charge distributions of secondary fragments produced by iron nuclei in polyethylene.
Our conclusions are formulated in Sec.~\ref{Conclusions}.

\section{Fermi break-up model for 
violent fragmentation of light nuclei}\label{FermiDescription}

For a light nucleus with a mass number $A_0$ and a charge $Z_0$ (in the 
following we assume $A_0\leq 16$) 
even a  relatively  small  excitation  energy  may  be  comparable
to its binding energy. In this case we assume  that  
the explosive decay of the excited nucleus into several smaller clusters is
the principal mechanism of de-excitation. To describe this
process we use a model which is similar to the famous 
Fermi break-up model for multiple particle production in proton-proton collisions~\cite{Fermi}. 
Later on it was extended to the multifragment break-up of highly excited nuclei,
see, e.g., Refs.~\cite{Gradsztajn,preprint90} and references therein. 
It is assumed that the excited nucleus decays simultaneously into cold or slightly 
excited fragments, which have lifetimes longer than the decay 
time, estimated as about 100 fm/c. The break-up configuration is characterized by 
some freeze-out volume $V$, where the produced fragments are placed.
The volume available for the translational motion, the so-called free volume $V_{f}$,  is smaller 
than the freeze-out volume, at least, by the proper volume of the 
fragments, $V_{f}=V-V_{0}$, where $V_0=A_0/\rho_{0}$ is the initial volume of the nucleus
at normal nuclear density $\rho_0=0.15$~fm$^{-3}$.
Below the following parameterization is used: $V_{f}=\kappa V_{0}$,
while $\kappa$ is a model parameter of order~1.   
The masses of fragments in their ground and lowest 
excited states were taken from nuclear data tables~\cite{Ajzenberg-Selove}.

We consider all possible break-up channels, which satisfy the mass number, charge, energy 
and momenta conservations, and take into account the competition between 
these channels. The central assumption of the model is that the probabilities of different 
break-up channels are proportional to their microcanonical weights.
In other words the probability of an individual break-up channel containing
$n$  particles  with  masses $m_{i}$ ($i=1,\cdots,n$) is proportional to its
phase space volume~\cite{Fermi,Gradsztajn,preprint90,Botvina87}:
\begin{equation} \label{eq:Fer}
W_{n}^{mic}\propto 
\frac{S}{G}\left(\frac{V_{f}}{(2\pi\hbar)^{3}}\right)^{n-1}
\left(\frac{\prod_{i=1}^{n}m_{i}}{M}\right)^{3/2}\frac{(2\pi)^
{\frac{3}{2}(n-1)}}{\Gamma\left(\frac{3}{2}(n-1)\right)}\cdot
\left(E_{kin}-U_{n}^{C}\right)^{\frac{3}{2}n-\frac{5}{2}},
\end{equation}
where $M=\sum_{i=1}^{n}m_{i}$ is the total mass of the fragments, 
$S=\prod_{i=1}^{n}(2s_{i}+1)$
is the spin degeneracy factor ($s_{i}$ is the $i$-th particle spin),
$G=\prod_{j=1}^{k}n_{j}!$
is the particle identity factor ($n_{j}$ is the number of particles of 
kind $j$). $E_{kin}$ is the total kinetic energy of  particles at
infinity which is related to the nucleus excitation energy 
$E_{A_{0}Z_{0}}^{*}$ as
\begin{equation} \label{eq:Ek}
E_{kin}=E_{A_{0}Z_{0}}^{*}+M_{0}c^2-\sum_{i=1}^{n}m_{i}c^2 .
\end{equation}
$M_{0}$ is mass of the decaying nucleus, 
$U_{n}^{C}$ is the Coulomb interaction energy between 
fragments given in the Wigner-Seitz approximation~\cite{Bondorf:etal:1995}: 
\begin{equation} \label{eq:UjC}
U_{n}^{C}=\frac{3}{5}\frac{e^{2}}{r_0}(1+\kappa)^{-1/3}
\left[\frac{Z_{0}^{2}}{A_0^{1/3}}-\sum_{i=1}^{n}\frac{{Z_i}^{2}}{{A_i}^{1/3}}\right],
\end{equation}
where $A_i$, $Z_i$ are mass numbers and charges of produced particles. 
In our calculations we have included 
fragments in all excited states, which are stable with respect to the
nucleon emission, as well as long-lived unstable nuclei 
$^{5}$He, $^{5}$Li, $^{8}$Be, $^{9}$B,
which decay at the later stage of the reaction~\cite{preprint90}. 

The number of channels included in our calculations was about $10^{3}$ for the $^{16}$O nucleus and 
$\sim 2\cdot 10^{2}$ for the $^{12}$C.
The Coulomb expansion stage was not considered explicitly 
for such light systems. The 
momentum distributions of final products were obtained
by the random generation over the whole accessible phase space, 
determined by the total kinetic energy, Eq.~(\ref{eq:Ek}), 
taking into account the energy and momentum
conservation. For the calculations of the available phase-space a very effective algorithm 
proposed by G.I.~Kopylov~\cite{KopylovNP1962,Kopylov} was used.

The Fermi break-up model in combination with the intranuclear cascade model
well describes various experimental data on 
disintegration of light nuclei by energetic protons with kinetic 
energies above 100 MeV~\cite{Bondorf:etal:1995,preprint90,Tashkent,sudov}.
However, as pointed out in Ref.~\cite{preprint90}, the model becomes less accurate below 100 MeV
when the emission channels with only few fragments dominate, in particular, in the case of 
$^{12}$C and $^{16}$O targets. 
The description of such two- and three-body decay channels can be improved
by considering alternative reaction mechanisms, e.g. 
the direct knock-out of light clusters, which do not requires the formation of a compound nucleus.

\section{Validation of Fermi break-up model of Geant4}\label{FermiValidation}

The Fermi break-model for decay of light nuclei was implemented as a FORTRAN-77 code,
see, e.g., Ref.~\cite{Bondorf:etal:1995,preprint90}. 
Later the same model was implemented in C++ by Vicente 
Lara~\cite{Geant4:Physics:Manual:G4FermiBreakUp,Lara:Wellisch:2000} and became 
a part of the Geant4 toolkit as the G4FermiBreakUp class.
The G4FermiBreakUp has a method BreakItUp which is applicable to  Geant4 objects of the type G4Fragment 
representing excited nuclei.  The BreakItUp results in a G4FragmentVector  which contains a set of 
G4Fragment objects. We can investigate now whether the results of the both implementations of the  
Fermi break-up model give consistent results.  Before running tests we have corrected 
the table of energy levels of excited light nuclei used by G4FermiBreakUp.

\subsection{Average multiplicities of nuclear fragments calculated with the Fermi break-up model}

\begin{figure}[htb]  
\begin{centering}
\includegraphics[width=1.1\columnwidth]{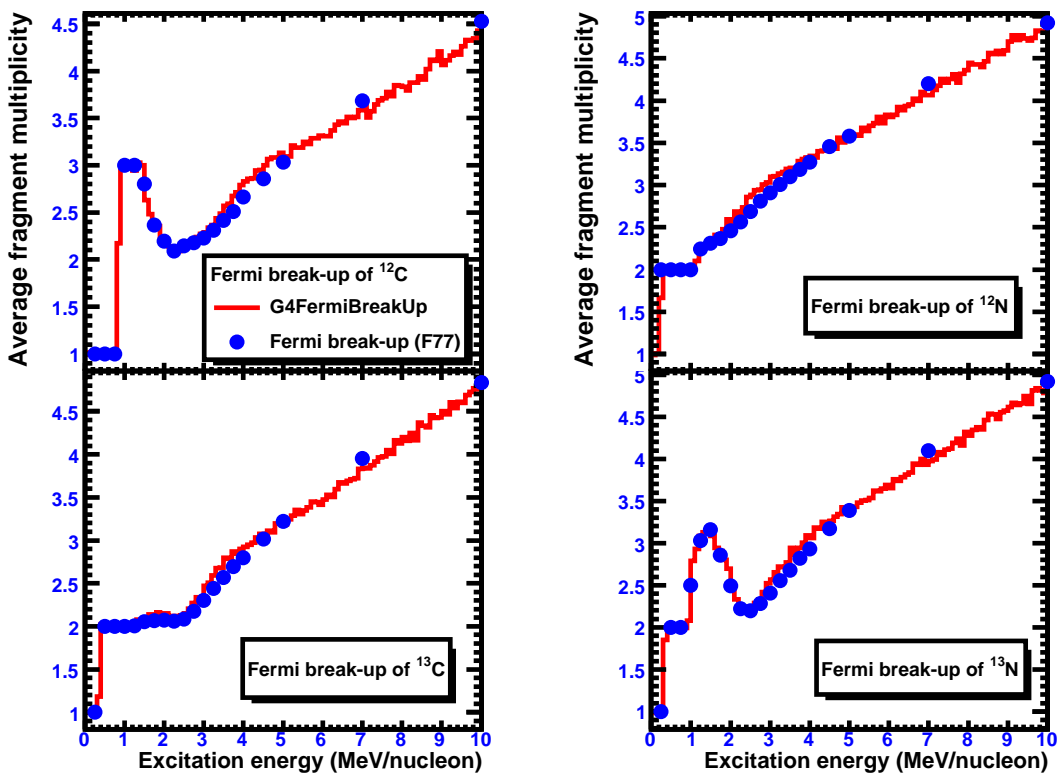}
\caption{Color online. Average multiplicities of nuclear fragments created in 
decays of excited $^{12}$C, $^{13}$C, $^{12}$N and $^{13}$N nuclei as functions of 
their excitation energies. The results of the G4FermiBreakUp of Geant4 are shown 
by histograms, the results of the Fermi break-up model of the SMM code are shown by points.}
\label{fig:multip_C12_C13_N12_N13}
\end{centering}
\end{figure}

The average multiplicity of nuclear fragments created in decays of excited nuclei is
an important characteristic of the fragmentation process which shows clearly its 
violent nature.
The average multiplicities calculated with the FORTRAN and C++ implementations of 
the Fermi break-up model are plotted in Fig.~\ref{fig:multip_C12_C13_N12_N13} 
as functions of excitation energy for decays of $^{12}$C, $^{13}$C, $^{12}$N and $^{13}$N.
Such highly-excited nuclei are abundantly produced, in particular, in interactions of 
therapeutic carbon-ion beams with tissue-like materials. These nuclei represent either 
excited projectile carbon nuclei or fragments of oxygen nuclei from the media. 
As seen from the figure, the predictions of two codes for 
average fragment multiplicities 
in decays of such nuclei are in very good agreement. 

\subsection{Average multiplicities of  hydrogen, helium, lithium and beryllium fragments}

\begin{figure}[htb]  
\begin{centering}
\includegraphics[width=1.1\columnwidth]{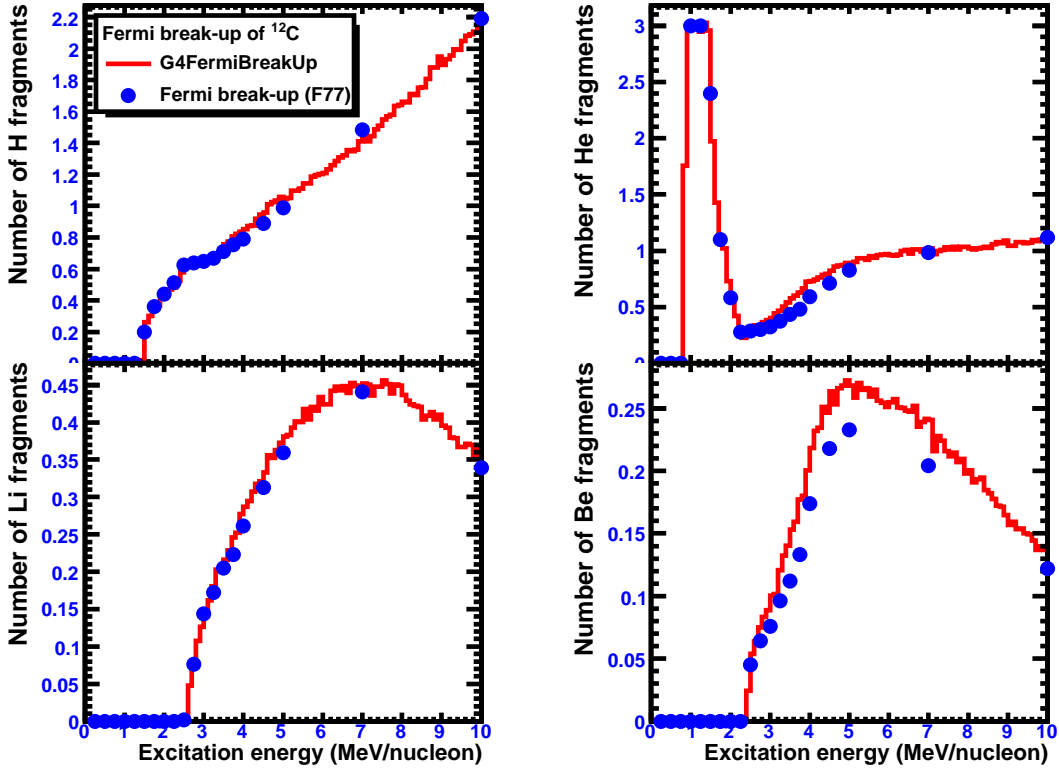}
\caption{Color online. Average multiplicities of H, He, Li and Be fragments created in 
decays of excited $^{12}$C as functions of its excitation energy.
The notations are the same as in Fig.~\ref{fig:multip_C12_C13_N12_N13}.}
\label{fig:C12_Z1_Z4}
\end{centering}
\end{figure}

The average multiplicities of  hydrogen, helium, lithium and beryllium fragments created in decays of 
$^{12}$C are shown in Fig.~\ref{fig:C12_Z1_Z4} as functions of $^{12}$C excitation energy.
One can see that the H and He isotopes are most abundant decay products at all excitation
energies. As seen in Figs.~\ref{fig:multip_C12_C13_N12_N13} and~\ref{fig:C12_Z1_Z4}, both implementations 
of the Fermi break-up model predict a peak at excitation energies of 1-2 MeV/nucleon corresponding to
the decay of $^{12}$C into three $\alpha$-particles. At excitations above 5-6 MeV/nucleon the 
nuclei decay mostly into protons and neutrons, but on average there is also 
one $\alpha$-particle among decay products. 
The production of Li and Be fragments reaches maximum at excitation energies 
of 7-8 and 5-6 MeV/nucleon, respectively. 
The differences between decays of
$^{12}$C, $^{13}$C, $^{12}$N and $^{13}$N nuclei reveal themselves only at low excitations,
e.g. below 2 MeV/nucleon. At 10 MeV/nucleon all nuclei decay into 4-5 fragments, on average, as
predicted by the both codes.

In general, the results obtained with two implementations of the Fermi break-up model agree well
at low and high excitations for all fragments, from H to Be. Some minor discrepancies in results 
are seen only for the yields of He and Be nuclei between 2.5 and 6 MeV/nucleon. 
We attribute these deviations to different nuclear mass tables used by the considered codes.

\section{Production of secondary fragments by light nuclei in tissue-like media}\label{YieldsC12}

\begin{figure}[htbp]  
\begin{centering}
\includegraphics[width=0.77\columnwidth]{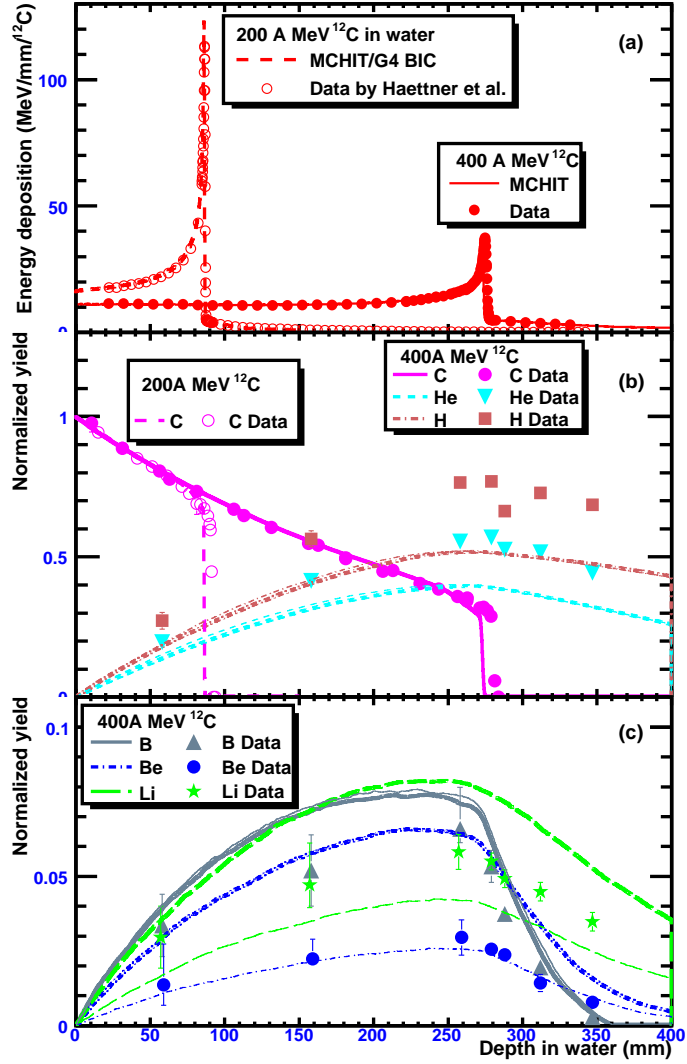}
\caption{Depth-dose distributions and normalized yields of 
secondary fragments for 200 and 400~$A$ MeV $^{12}$C in water calculated with MCHIT involving
the Light-Ion Binary cascade model:
(a) the average linear energy deposition per beam particle;
the yields of nuclear fragments per beam particle (b) for carbon, helium and hydrogen fragments;
(c) for boron, beryllium and lithium fragments. The yields of secondary fragments calculated 
with G4FermiBreakUp applied after the cascade model are shown by thick lines.
The same yields, but calculated with the evaporation model applied after the Light-Ion 
Binary cascade model
are shown by thin lines. 
Experimental data from Ref.~\cite{Haettner:Iwase:Schardt:2006} are shown by various symbols.}
\label{fig:C12_200_400_Water_dose_Z_9_1_BIC}
\end{centering}
\end{figure}

\begin{figure}[htbp]  
\begin{centering}
\includegraphics[width=0.77\columnwidth]{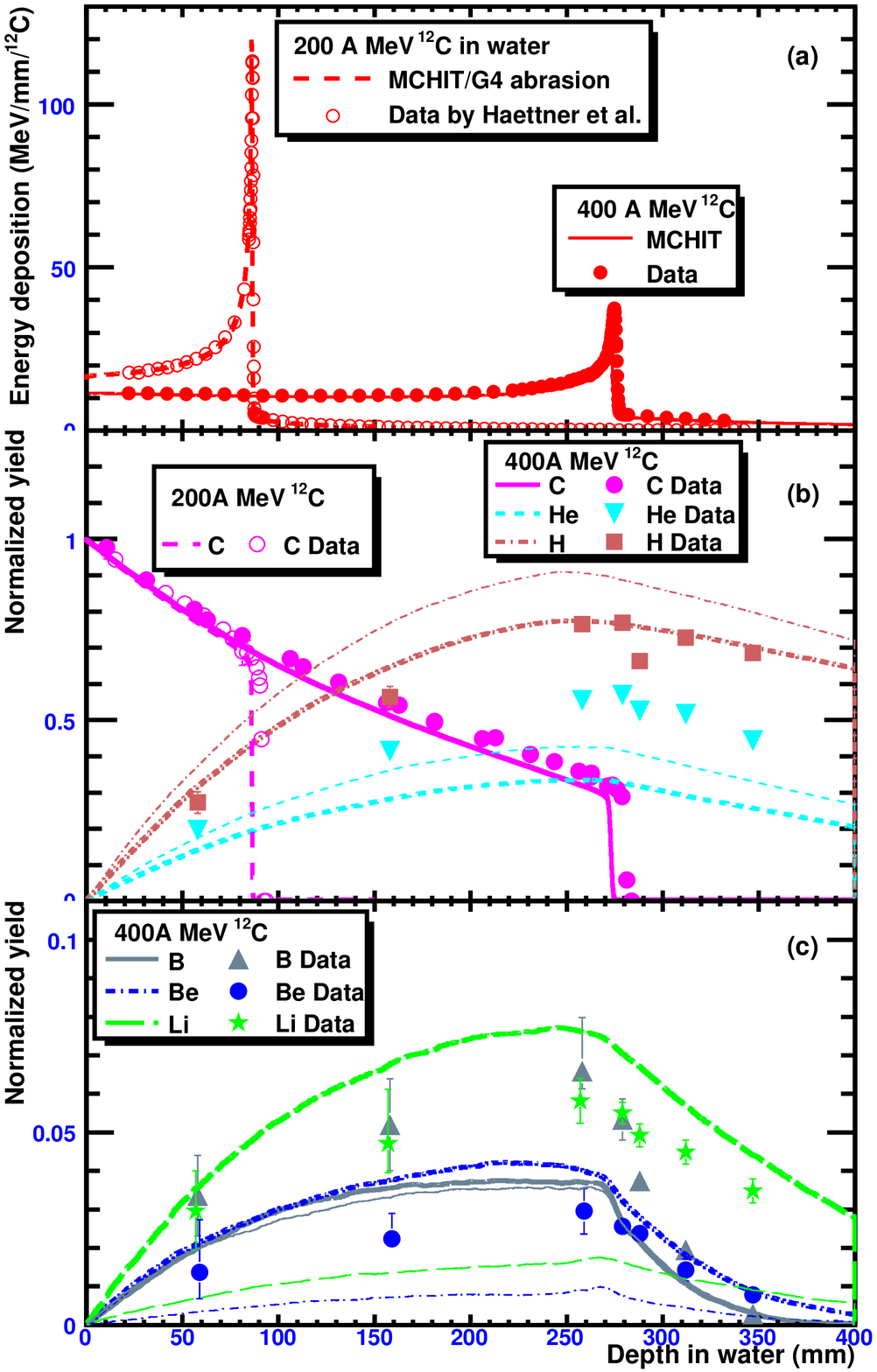}
\caption{Depth-dose distributions and normalized yields of 
secondary fragments for 200 and 400~$A$ MeV $^{12}$C in water calculated with MCHIT involving
the Wilson abrasion model. The notations are the same as in 
Fig.~\ref{fig:C12_200_400_Water_dose_Z_9_1_BIC}.}
\label{fig:C12_200_400_Water_dose_Z_9_1_abr}
\end{centering}
\end{figure}

The depth-profiles of average linear energy deposition calculated per beam particle 
for 200 and 400~$A$ MeV $^{12}$C in water are 
shown in Figs.~\ref{fig:C12_200_400_Water_dose_Z_9_1_BIC} and 
\ref{fig:C12_200_400_Water_dose_Z_9_1_abr}. They were obtained with the Light-Ion Binary cascade model 
and the Wilson abrasion model, respectively.
The calculations were performed for a water phantom of $40\times 15\times 15$ cm, 
which was divided into 0.25~mm slabs. The energy deposition was calculated for each of the slabs.
At the end of the run it was divided by the slab thickness and by the number of projectile nuclei 
to estimate the average linear energy transfer per beam particle. 
A Gaussian beam cross section of 5 mm FWHM was assumed. The beam energy distribution was 
taken as a Gaussian with the FWHM of 0.2\% of the mean energy.

Experimental data of Ref.\cite{Haettner:Iwase:Schardt:2006} for relative ionization in water for 
200 and 400~$A$ MeV $^{12}$C  beams were rescaled to the calculated absolute dose values at zero depth.
These data are plotted in Figs.~\ref{fig:C12_200_400_Water_dose_Z_9_1_BIC} and
\ref{fig:C12_200_400_Water_dose_Z_9_1_abr} and compared to the calculations using both
nucleus-nucleus collision models mentioned above.  
In both cases the MCHIT model successfully describes the positions of the Bragg peak 
and the peak-to-plateau ratios of the depth-dose curves for 200 and 400~$A$ MeV $^{12}$C ions in water. 
We used the set of standard electromagnetic models of Geant4 for calculating ionization 
energy losses of carbon nuclei and their fragments. For better description of the 
positions of the Bragg peaks the average ionization potential for water molecule was set to 81 eV,
which is in the range of uncertainty quoted for this quantity~\cite{Paul:2007}. 
  
The normalized yields of primary carbon nuclei and secondary nuclear fragments 
(hydrogen, helium, lithium, beryllium and boron nuclei)
were also calculated with MCHIT and plotted in 
Figs.~\ref{fig:C12_200_400_Water_dose_Z_9_1_BIC} and ~\ref{fig:C12_200_400_Water_dose_Z_9_1_abr}.  
In these figures the calculated yields are compared with the 
measurements of Ref.~\cite{Haettner:Iwase:Schardt:2006}.
As the yields were measured by detecting fragments
within a $10^{\rm o}$ cone surrounding the beam axis, 
as explained in Ref.~\cite{Haettner:Iwase:Schardt:2006},
corresponding acceptance cuts were introduced in calculations.

Both the Light-Ion Binary cascade and  Wilson abrasion models describe rather well the observed 
attenuation of the carbon beam in water at 200 and 400~$A$ MeV. The measured and calculated 
carbon yields correspond to the mixture of primary $^{12}$C and secondary carbon nuclei like 
$^{9}$C, $^{10}$C and $^{11}$C. The last two nuclides ($^{10}$C  and $^{11}$C) 
are relevant to PET monitoring of dose distribution in carbon-ion therapy.
In Ref.~\cite{Pshenichnov:etal:2006} their yields were calculated with the MCHIT model 
and compared with available experimental data. As follows from the data and calculations, 
{\em more} than 40\% of primary 200~$A$ MeV $^{12}$C nuclei undergo fragmentation on their 
way to the Bragg peak, while at 400~$A$ MeV this fraction 
exceeds 70\%. 

The production of H, He, Li, Be and B nuclei is characterized by
gradually increasing yields until primary ions stop at the depth, 
where the Bragg peak is located. Secondary fragments propagate further beyond the distal 
edge of the Bragg peak, and their yields decrease due to secondary 
fragmentation reactions on nuclei in the media. 
The boron yield drops fastest since B has the largest charge and inelastic cross section 
among the elements produced by $^{12}$C ions. On the contrary, H and He 
fragments are less attenuated and most of them leave the water phantom of  
40~cm depth.

By inspecting Figs.~\ref{fig:C12_200_400_Water_dose_Z_9_1_BIC} 
and~\ref{fig:C12_200_400_Water_dose_Z_9_1_abr} one can compare the Light-Ion Binary cascade model and
the Wilson abrasion model with respect to their ability to describe secondary fragments.
The yields of secondary fragments calculated with the evaporation model instead of 
the Fermi break-up model are also shown for comparison. 

The cascade model underestimates the yields of hydrogen fragments, while the abrasion model
successfully describes them when the Fermi break-up model is used instead of the evaporation model.
Both nucleus-nucleus collision models underestimate He yields, and the inclusion of G4FermiBreakUp
does not improve this situation. This deficiency is apparently caused by neglecting the  
alpha-clustered structure of $^{12}$C, which may enhance the production of $\alpha$-particles 
(see also Ref.~\cite{Tashkent}).

As seen from Fig.~\ref{fig:C12_200_400_Water_dose_Z_9_1_BIC}, the cascade model reasonably describes
the B yields, but fails to reproduce the production of Li and Be.  
When the evaporation model is used instead of G4FermiBreakUp, 
the agreement with Be data is achieved, but the production of Li is strongly underestimated. 
The abrasion model gives more accurate predictions for the yields of Li and Be fragments,
but only in combination with the Fermi break-up model. When instead the evaporation model is used,
the Li and Be yields are underestimated by large factors. At the same time both de-excitation models
underestimate the B yields. 
This can be explained either by the failure of the abrasion model to describe 
peripheral nucleus-nucleus collisions, leading to low-excited residual nuclei, 
or by the inaccuracy of the Fermi break-up model in describing the two-body 
decays of nuclei with lowest excitations.   

In summary, our analysis shows that no one of the considered models describes the production 
of secondary fragments perfectly. Generally, the fragment yields are somewhat better 
described by the abrasion model as compared with the cascade model. However, the former 
model still underestimates the yields of 
He and B fragments.  The inclusion of the Fermi break-up model at the de-excitation stage of nuclear 
reaction improves the description of H, Li and Be yields, but does not help to remove 
discrepancies for B fragments. This is an important issue for Monte Carlo calculations for carbon-ion therapy. 
Indeed, as shown recently~\cite{Nose:et:al:2009}, the dose and radiation quality at the center of 
a broad carbon-ion beam formed by using a passive beam delivery system are influenced by  
such secondary fragments from the off-center region.

\section{Break-up of medium-weight and heavy nuclei simulated with SMM model}\label{SMMDescription}

The Statistical Multifragmentation Model (SMM) is based on the assumption of
statistical equilibrium between produced fragments in a
low-density freeze-out volume~\cite{Bondorf:etal:1995}. 
We believe that the chemical composition, i.e. masses and charges of primary fragments, 
are fixed at this stage.
However, the fragments can still interact with other nuclear
species via the Coulomb and nuclear mean fields. Hence their
energies and densities may be affected by these residual
interactions. 

An advantage of the model is that all breakup channels to nucleons and
excited fragments are considered within the same statistical framework, 
and, in particular, the formation of a compound 
nucleus is included as one of the channels. This allows for a smooth transition from
the decay via evaporation and fission at low excitation energies~\cite{Bohr} 
to the multifragmentation at high excitations. 

In the microcanonical treatment~\cite{Bondorf:etal:1995,Botvina01} 
the mass, charge, momentum and energy of the system are strictly fixed.
It is also assumed that the primary fragments are formed 
in the expanded volume $V > V_0$, where $V_0$ is the volume at normal nuclear density
$\rho_0=0.15$ fm$^{-3}$. 

In accordance with the statistical hypothesis the probability of the decay channel $j$
is given by the statistical weight $W_j \propto \exp{S_j}$, 
where $S_j$ is the entropy of the system in
channel $j$ which is a function of the excitation energy $E_x$, 
mass number $A_{0}$, charge $Z_{0}$ and other global
parameters of the source. After formation in the freeze-out
volume, the fragments propagate independently in their mutual
Coulomb field and undergo secondary decays. De-excitation of
the hot primary fragments proceeds via evaporation, fission, or
Fermi-breakup~\cite{Botvina87}.

In the SMM light fragments with mass number $A\le 4$ and charge
$Z\le 2$ are considered as structureless particles (nuclear gas)
with their masses and spins taken from the nuclear data tables. Only
translational degrees of freedom of these particles contribute to
the entropy of the system. Fragments with $A > 4$ are treated as
heated drops of nuclear liquid, and their individual free energies
$F_{AZ}$ are parameterized according to the liquid drop model:
\begin{equation}
F_{AZ}=F^{B}_{AZ}+F^{S}_{AZ}+E^{C}_{AZ}+E^{Sym}_{AZ},\label{FreeEnergy}
\end{equation}
where terms in the r.h.s correspond to the bulk, surface, Coulomb
and symmetry energy.
In this expression $F^{B}_{AZ}=(-W_0-T^2/\epsilon_0)A$ is
the bulk energy term including the contribution of internal
excitations controlled by the level-density parameter
$\epsilon_0$, and $W_0 = 16$~MeV is the binding energy of infinite
nuclear matter.
$F^{S}_{AZ}=B_0A^{2/3}((T^2_c-T^2)/(T^2_c+T^2))^{5/4}$ is the
surface energy term, where $B_0=18$~MeV is the surface coefficient
at $T=0$, and $T_c=18$~MeV is the critical temperature of infinite
nuclear matter. The Coulomb energy of individual fragments is
calculated as $E^{C}_{AZ}=\frac{3}{5}\frac{Z^2 e^2}{r_0 A^{1/3}}c(\rho)$,
where $e$ is the proton charge, $r_0$=1.17 fm, and the last factor,
$c(\rho)=1-(\rho_{p}/\rho_{p0})^{1/3}$, where $\rho_{p0}$ is the normal proton
density of nuclei $\rho_{p0}\approx \frac{Z}{N}\rho_0$. This factor
describes the screening effect due to 
the presence of other fragments in the Wigner-Seitz approximation.
The last term in Eq.~(\ref{FreeEnergy}) $E^{Sym}_{AZ}=\gamma (A-2Z)^2/A$ 
is the symmetry energy term,
where $\gamma = 25$~MeV is the symmetry energy coefficient. These
parameters are taken from the Bethe-Weizs\"acker formula and
correspond to the isolated cold fragments with normal nuclear density.
This assumption has been proven to be quite successful in many
applications. However, a more realistic treatment of primary fragments
in the freeze-out volume may require certain modifications of the
liquid-drop parameters as suggested by experimental 
data~\cite{LeFevre,Ogul:et:al:2006,Souliotis:et:al:2007}.

In the grand canonical (macrocanonical) version of the SMM~\cite{Botvina85},
after integrating out translational degrees of freedom, one can
write the mean multiplicity of nuclear fragments with mass $A$ and charge $Z$
as
\begin{eqnarray}
\label{naz} \langle N_{AZ} \rangle =
g_{AZ}\frac{V_{f}}{\lambda_T^3}A^{3/2} {\rm
exp}\left[-\frac{1}{T}\left(F_{AZ}(T,\rho)-\mu A-\nu
Z\right)\right]. \nonumber
\end{eqnarray}
Here $g_{AZ}$ is the ground-state degeneracy factor of species
$(A,Z)$, $\lambda_T=\left(2\pi\hbar^2/m_NT\right)^{1/2}$ is the
nucleon thermal wavelength, and $m_N \approx 939$ MeV is the 
nucleon mass. Here
$V_f\approx \kappa V_0$ is the free volume available for the
translational motion of fragments and $\kappa$ is the model parameter which
in principle can depend on the fragment 
multiplicity in the freeze-out volume~\cite{Bondorf:etal:1995}.
The chemical potentials $\mu$ and $\nu$ are found from the mass and charge constraints:
\begin{equation} \label{eq:ma2}
\sum_{(A,Z)}\langle N_{AZ}\rangle A=A_{0},~~ \sum_{(A,Z)}\langle
N_{AZ}\rangle Z=Z_{0}.
\end{equation}

Numerous comparisons of the SMM calculations with
experimental data on thin targets show that generally the model describes data very well (see,
e.g., Refs.~\cite{Botvina90,ALADIN,EOS,MSU,INDRA,FASA,Dag}). This
demonstrates that the statistical approach with liquid-drop
description of individual fragments provides adequate treatment of
the multifragmentation process.

\section{Validation of statistical multifragmentation model of Geant4}\label{SMMValidation}

\subsection{General remarks}

The statistical approach to multifragment break-up of hot nuclear
systems outlined above was first formulated in Refs.~\cite{Botvina85,Bondorf:1985},
and it is well-known now as a Copenhagen-Moscow model 
(for a review see Ref.~\cite{Bondorf:etal:1995}).  
For numerical calculations it was initially implemented as a FORTRAN-77 SMM code,
see, e.g., Ref.~\cite{Botvina87}. 
Later the SMM was implemented in C++ by Vicente 
Lara~\cite{Lara:Wellisch:2000,Geant4:Physics:Manual:G4StatMF} and became 
a part of the Geant4 toolkit as the G4StatMF class.
G4StatMF has a method BreakItUp which is applicable to a Geant4 object of type G4Fragment 
representing a nucleus in its ground or excited state. 
After application of this method a G4FragmentVector is produced as an output, which 
consists of a set of G4Fragment objects. 

It is worthwhile to mention that the C++ implementation of the SMM
has been developed following the FORTRAN version of the model~\cite{Botvina87}. 
After inspecting the source code of G4StatMF we came to the conclusion
that the physical parameters of the model are basically the same as in
Refs.~\cite{Bondorf:etal:1995,Botvina01}.
However, the numerical methods involved to find, e.g. the temperature and 
chemical potentials for a macrocanonical ensemble of nuclear fragments,  
are different in the FORTRAN and C++ implementations of the SMM. The FORTRAN SMM works with
single precision floating point numbers, while the G4StatMF uses double precision floats.
In view of these differences we performed a systematic comparison of numerical results 
delivered by the two codes.

The G4StatMF from the Geant4 toolkit of version 9.1 was used to simulate the multifragment break-up
of hot nuclear systems with mass and charge of $^{112}$Sn and $^{208}$Pb. 
In stand-alone tests  $10^5$ decay events were generated for each nuclear system and excitation 
energy of 3, 4, 5 and 8 MeV/nucleon. 
In these calculations the parameter $\kappa$ which defines the free volume 
for translational motion of fragments was taken as a 
function of each event's fragment multiplicity as $\kappa (M_f)$, where $M_f=\sum_{i}N_i$.
During the simulation several kinds of 
distributions of produced fragments were scored. Such distributions were compared with the
results obtained with the FORTRAN SMM using the same calculational 
parameters.

\subsection{Fragment mass distributions}

At the beginning of our study severe discrepancies between the codes were found
for the mass distributions of produced fragments. The disagreements were already seen for the decay 
of $^{112}$Sn, but they became more pronounced when we considered nuclear systems 
far from the stability line, e.g. $^{55}$Sn and $^{140}$Sn.
Such exotic proton- and neutron-rich systems cannot be produced in
collisions of stable nuclei, but we consider them only for the verification of the code.  
Our simulations of multifragment break-up of such exotic nuclear systems 
lead us to the conclusion that the calculation of the fragments' symmetry energy in 
G4StatMF of version 9.1 was wrong. The error was localized and corrected.
After that the results of the two implementations of the multifragmentation model
turned out to be very close. For example, the ensemble's average 
temperatures  in decays of $^{112}$Sn and $^{140}$Sn 
differed by less than 0.2 MeV instead of 1 MeV before the corrections.
In addition, several changes were introduced to improve the stability of G4StatMF 
when applied to proton- and neutron-rich systems like 
$^{23}$O, $^{55}$Sn and $^{50}$Ne, $^{140}$Sn.
After these improvements to the G4StatMF were introduced, the corresponding 
distributions were calculated again in order to see the effect of updates. 

\begin{figure}[htb]  
\begin{centering}
\includegraphics[width=1.1\columnwidth]{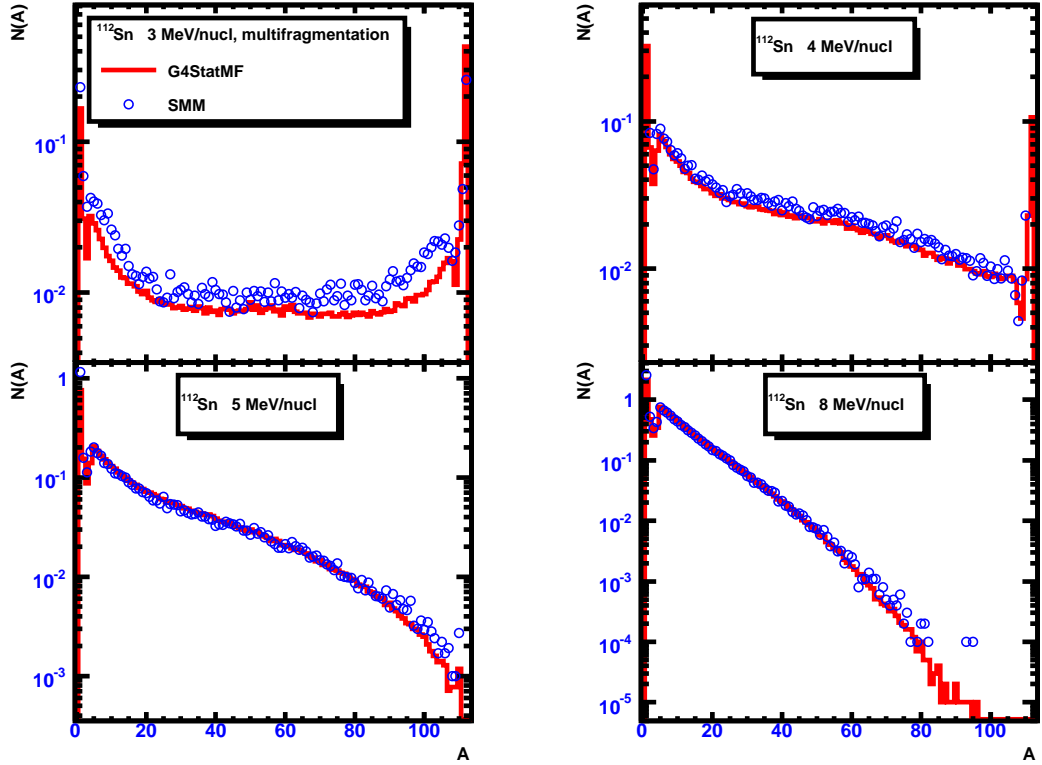}
\caption{Color online. Mass distributions of nuclear fragments created after the 
decay of excited $^{112}$Sn with excitation energies of
3, 4, 5 and 8 MeV/nucleon. The results of the modified G4StatMF  of Geant4 are given 
by histograms, the results of the FORTRAN SMM  - by open circles.}
\label{fig:massDistSn112}
\end{centering}
\end{figure}

The results for $^{112}$Sn are shown in Fig.~\ref{fig:massDistSn112} for  
excitation energies of 3 and 4 MeV/nucleon  where competition of explosive 
multifragment break-up and evaporation from the compound nucleus is taking place. 
This is clearly reflected in the mass distributions by the presence 
of heavy evaporation residues with $A\sim 110$. 
However, as clearly seen at 3 and 4 MeV/nucleon, the G4StatMF predicts much stronger compound 
nucleus peak than the standard SMM. Apparently this can be explained by a larger fraction 
of evaporation events produced by G4StatMF. 
At higher excitation energies the mass distributions produced with both 
codes agree quite well. At 5 and 8 MeV/nucleon the peaks associated with 
evaporation residues completely disappear and 
multifragment break-up becomes a dominating channel. It is seen also, that
the production of intermediate mass fragments (IMF) with $4<A<20$ is enhanced.
At such high excitations the results of both codes are in perfect agreement both in shape and 
in absolute values.    

\begin{figure}[htb]  
\begin{centering}
\includegraphics[width=1.1\columnwidth]{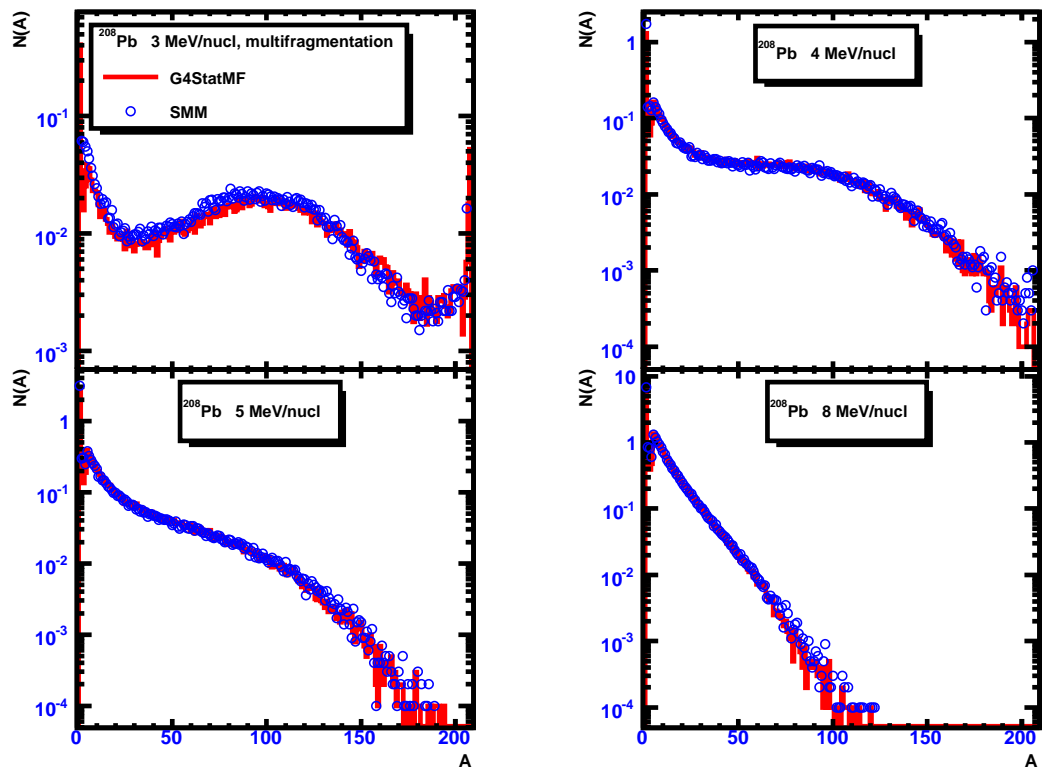}
\caption{Color online. Same as in Fig.~\ref{fig:massDistSn112} but for the 
decay of excited $^{208}$Pb nucleus.}
\label{fig:massDistPb208}
\end{centering}
\end{figure}

In Fig.~\ref{fig:massDistPb208} the results for a heavy system, $^{208}$Pb, are shown for
excitation energies of 3, 4, 5 and 8 MeV/nucleon.  At 3 MeV/nucleon the contribution of 
fission-like events is clearly visible. Such events are characterized by creation of 
two fragments with comparable mass numbers of $A\sim 100$. Both SMM implementations 
predict such a behavior. At higher excitations the peak from symmetric fission-like 
events disappears. As seen from Fig.~\ref{fig:massDistPb208}, 
the mass distributions of fragments calculated with G4StatMF and FORTRAN SMM are in 
good agreement both in shape and in absolute values.  

As additional test for the consistency of the two SMM implementations 
one can  calculate the average charge of the fragment as a function of the fragment mass,
$\langle Z \rangle (A)$. 
It is expected that for each fragment mass the ratio $\langle Z \rangle/A$
is close to the ratio for the initial hot system. We found that the results of 
G4StatMF and the FORTRAN version of SMM follow this trend. Since the results of 
the two codes completely coincide, we do not present the corresponding plots here.

\subsection{Fragment charge distributions}

Further validation of G4StatMF is possible with experimental data on low-energy 
(several tens of MeV/nucleon) nucleus-nucleus collisions. This energy regime allows 
colliding nuclei to overcome the Coulomb barrier and create a single combined system.
There exist dedicated experimental techniques~\cite{Rivet:etal:1998,Tabacaru:et:al:2003},
which make possible to select central collision events, resulting in a single emitting 
source, and allow to estimate the source's charge, mass and excitation energy. 

\begin{figure}[htb]  
\begin{centering}
\includegraphics[width=1.05\columnwidth]{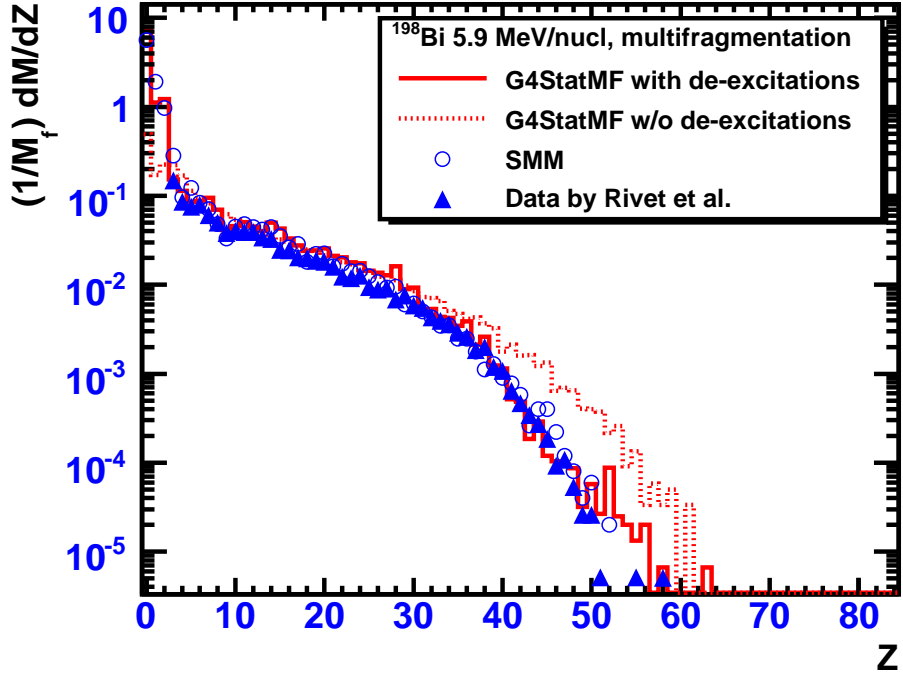}
\caption{Color online. Fragment charge distributions (normalized to each event's 
$Z\ge 5$ multiplicity) in decay of hot $^{198}$Bi-like system 
representing the equilibrated source formed in central $^{129}$Xe+$^{\rm nat}$Sn collisions. 
The results of modified G4StatMF with and without accounting for secondary de-excitations of produced
fragments are given by solid- and dashed-line histograms, respectively. 
The results of the FORTRAN SMM are presented by open circles. 
Triangles represent experimental data~\cite{Rivet:etal:1998}.
}
\label{fig:ChargeDistRivetXeSn}
\end{centering} 
\end{figure}

In particular, differential charge multiplicity distributions were measured in 
Ref.~\cite{Rivet:etal:1998} for central $^{129}$Xe+$^{\rm nat}$Sn collisions at 32 $A$ MeV.
The event selection procedure adopted in Ref.~\cite{Rivet:etal:1998} required detection of a
significant fraction (80\%) of the total charge and of the initial linear momentum of colliding nuclei.
Therefore, one can estimate the mass and charge of the emitting source as $A=198$ 
and $Z=83$ roughly corresponding to a very hot $^{198}$Bi nucleus. Events containing less than
three fragments with charge $Z \ge 5$ were rejected to favor events  
resulting from the multifragmentation process.  Differential charge multiplicity 
distributions  for decay of $^{198}$Bi were
calculated with G4StatMF and plotted in Fig.~\ref{fig:ChargeDistRivetXeSn}.
We have found that the experimental data are best described under the assumption that 
the  $^{198}$Bi system has the excitation energy of 5.9 MeV/nucleon. This value is lower compared to
the available excitation energy per nucleon for a hypothetic compound system formed
via a complete fusion of Xe and Sn nuclei~\cite{Rivet:etal:1998}.
This difference can be explained by the fact that the most energetic nucleons leave the system 
before the statistical equilibrium is reached, so that 
the average excitation energy per nucleon drops down.

The fragment charge distributions were filled according to the event selection procedures 
adopted in the experiment~\cite{Rivet:etal:1998}. In particular, following simulation of each
individual event, the multiplicity of fragments with $Z\ge 5$ in the event was calculated.
Then the numbers of fragments of certain charge in the event were calculated and divided by 
this multiplicity value. After that the corresponding histogram shown in Fig.~\ref{fig:ChargeDistRivetXeSn}
was filled with these numbers.

Distribution of hot fragments produced in multifragmentation events are shown by the dotted 
histogram in Fig.~\ref{fig:ChargeDistRivetXeSn}. As one can see, they significantly
differ from the experimental data. We have found that the agreement with data is improved 
significantly if we account for secondary de-excitation of these fragments.
This is especially important for heavy fragments with $Z > 30$. 
The final distributions calculated with G4StaMF agree well
with the experimental data of Ref.~\cite{Rivet:etal:1998}, as well as with the distributions
calculated with the FORTRAN SMM.

\subsection{IMF multiplicity distributions}

\begin{figure}[htb]  
\begin{centering}
\includegraphics[width=1.05\columnwidth]{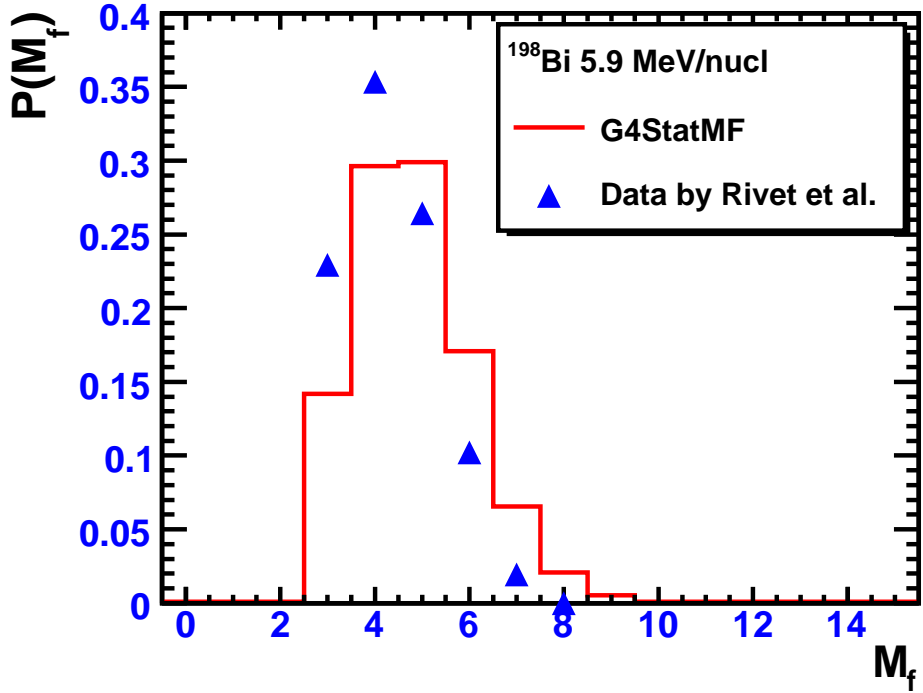}
\caption{Color online. Multiplicity distributions of intermediate mass   
($Z\ge 5$) fragments in decay of hot $^{198}$Bi-like system 
representing the equilibrated source formed in central $^{129}$Xe+$^{\rm nat}$Sn collisions.
The results of modified G4StatMF are given 
by histograms. Triangles represent experimental data~\cite{Rivet:etal:1998}.
}
\label{fig:MfDistRivetXeSn}
\end{centering} 
\end{figure}

Now one can calculate the multiplicity distributions of intermediate mass fragments 
(IMF, $4 < A < 20$) produced at the break-up of
a  $^{198}$Bi-like nucleus  created in central $^{129}$Xe+$^{\rm nat}$Sn collisions.  
Taking into account secondary decays the average IMF multiplicity calculated with the G4StatMF is 4.8. 
This value is slightly higher than the  experimental value  of 4.3 
reported in Ref.~\cite{Rivet:etal:1998}. 
The IMF multiplicity distribution calculated with G4StatMF is  shown in 
Fig.~\ref{fig:MfDistRivetXeSn} together with the distribution measured in Ref.~\cite{Rivet:etal:1998}. 

As seen from Fig.~\ref{fig:MfDistRivetXeSn}, the general shape of the experimental 
distribution is well reproduced by G4StatMF, but the model distribution is shifted
towards higher multiplicities.  We attribute this deficiency to the uncertainties in 
the choice of the system size, which is smaller than the $^{198}$Bi-source used in the simulations.
As shown in Ref.~\cite{Rivet:etal:1998}, the IMF multiplicity distributions are 
rather sensitive to the size of the emitting source.

At the same time, the  uncertainties in system size do not affect very much the fragment charge 
distributions shown in 
Fig.~\ref{fig:ChargeDistRivetXeSn}, where the experimental data are described well even without the 
exact specification of the system size in calculations.  
Indeed, as pointed out in Ref.~\cite{Rivet:etal:1998}, 
the fragment charge distributions for hot nuclear systems are insensitive to the system size 
but are mostly governed by the system's excitation energy per nucleon.

\section{Role of multifragmentation in
transport calculations of medium-weight nuclei in extended media}\label{IronInWater}

After the validity of the G4StatMF has been confirmed, we can 
estimate the influence of multifragment break-up on ion transport in tissue-like materials. 
It is evident that the yields of secondary fragments, especially of IMFs
$4 < A < 20$,  produced by beams of heavy nuclei in thick targets depend on the model
used to simulate de-excitation of residual nuclei. Before considering fragment yields,
we examine more rough characteristic of ion transport through tissue-like materials, namely,  
the depth-dose distributions in water.

\subsection{Depth-dose distributions of $^{56}$Fe beams in water}\label{FeDepthDose}

The MCHIT model~\cite{MCHIT} described above has been extensively used to study ion propagation in 
tissue-like materials. Several examples of the depth-dose distributions relevant to ion-beam cancer therapy 
can be found in Refs.~\cite{Pshenichnov:etal:2005,Pshenichnov:etal:2006,Pshenichnov:etal:2007,MCHIT}.
Here we present results for heavier incident nuclei, Fe, which are relevant to the evaluation of cosmic
radiation effects.

\begin{figure}[htbp]  
\begin{centering}
\includegraphics[width=0.95\columnwidth]{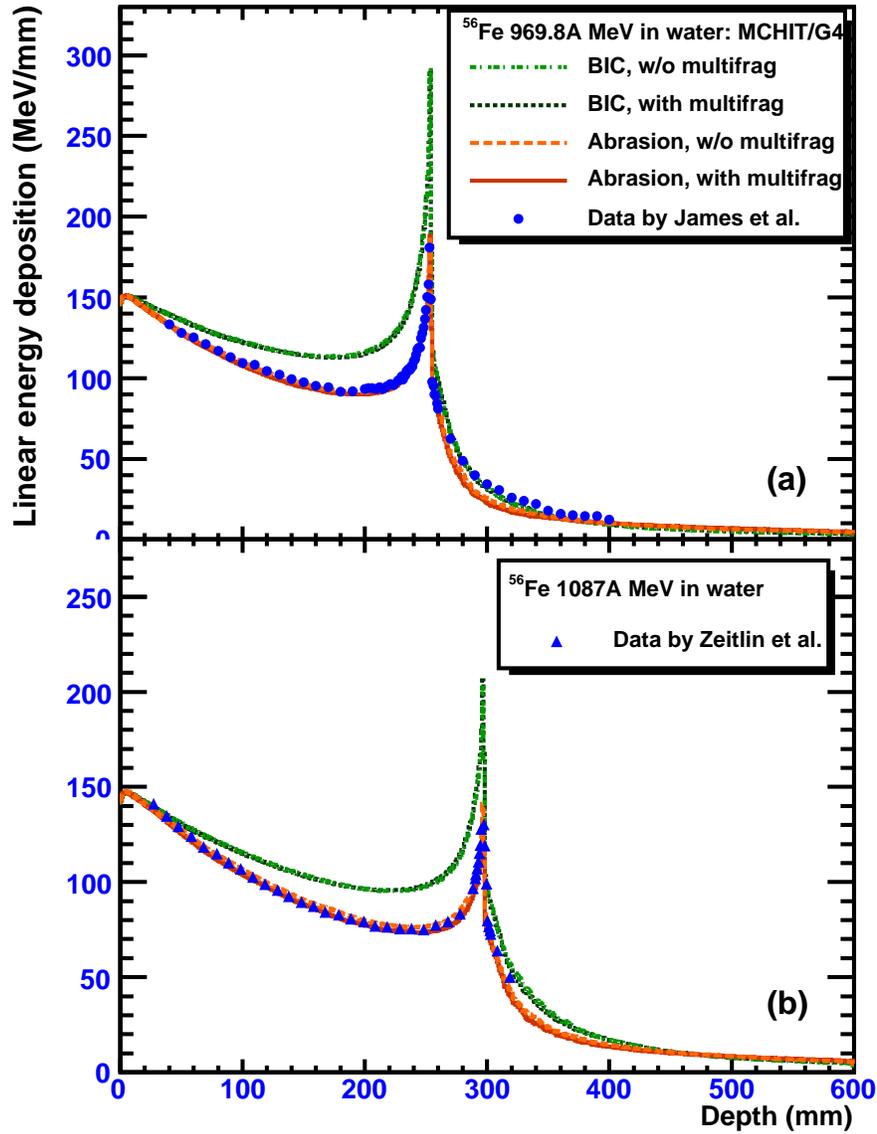}
\caption{Color online. Depth-dose distributions of (a) 969.8 $A$ MeV and 
(b) 1087 $A$ MeV $^{56}$Fe nuclei in water
calculated with the MCHIT. The results of the Light-Ion Binary cascade  
model of Geant4 with and without multifragmentation are shown by the dash-dotted and dotted 
histograms, respectively. The results of the Wilson abrasion model with and 
without multifragmentation are shown by the solid and dashed
histograms, respectively. The experimental data from 
Refs.~\cite{James-Moyers:etal:2006,Zeitlin:Heilbronn:Miller:1998} are shown by various points.}
\label{fig:Fe_969_1087_Water}
\end{centering}
\end{figure}

The depth-dose distribution for 969.8 $A$ MeV and 1087 $A$ MeV  $^{56}$Fe nuclei 
in water calculated with MCHIT are shown in Fig.~\ref{fig:Fe_969_1087_Water}
together with the experimental data of 
Refs.~\cite{James-Moyers:etal:2006} and~\cite{Zeitlin:Heilbronn:Miller:1998}, respectively.  
A water phantom of $60\times 10\times 10$ cm was divided into 0.5~mm slabs,
and the average linear energy deposition was calculated in each of the slabs.  
The calculations were performed for a Gaussian beam profile of 4 mm FWHM in the transverse plane to the beam
direction. The energy spread of the beam was assumed to be Gaussian with 
the FWHM of 0.2\% of the reported mean beam energy.
Similar to calculations presented in Sec.~\ref{YieldsC12}, two models 
describing the initial stage of inelastic nucleus-nucleus collisions, 
the Light-Ion Binary cascade model~\cite{Geant4:Physics:Manual:BIC} and the 
Wilson abrasion model~\cite{QinetiQ:Manual,Geant4:Physics:Manual:Abr} were used in independent runs. 
 
De-excitation of hot nuclear residues formed in collisions of Fe ions with nuclei of the medium
after emission of fast nucleons was treated by 
using the BreakItUp method of the class G4ExcitationHandler. 
Within the G4ExcitationHandler several de-excitation models can be applied, namely,
nucleon evaporation, photon emission, Fermi break-up and nuclear 
multifragmentation. The G4ExcitationHandler has public methods which set the domains of
applicability of the de-excitation models in terms of the size and excitation energy 
of nuclear residues. However, since the Light-Ion Binary cascade model
creates its own private instance of G4ExcitationHandler, such methods are currently 
not available to the user when this model is involved.
By default, the multifragmentation is switched off, and we have changed the source code of the
G4ExcitationHandler to activate it. On the contrary, The G4WilsonAbrasionModel has public methods 
which control the behavior of the G4ExcitationHandler.  
The distributions presented in Fig.~\ref{fig:Fe_969_1087_Water} 
were calculated both with and without accounting for multifragment break-up at excitation energies 
$> 3$ MeV/nucleon. 

The probability of nuclear interactions for high-energy Fe nuclei in water is rather high.
One can estimate the mean free path of the beam with respect to nuclear interactions,
$\lambda_{\rm nuc}=1/(\sigma_{\rm nuc} n_{\rm water(O)})\sim 17$~cm, 
where $\sigma_{\rm nuc}\sim 2$~barn is the total nuclear
reaction cross section in Fe+O collisions and $n_{\rm water(O)}=3\times 10^{22}$~cm$^{-3}$ is the 
density of oxygen nuclei in water.

Nuclear interactions are reflected in the shape of the depth dose-distribution presented in 
Fig.~\ref{fig:Fe_969_1087_Water}. A globally descending trend is explained by the destruction of beam
particles in fragmentation reactions on nuclei of the medium. These reactions lead not only to
reduction of the Bragg peak, but also to  build-up of a tail beyond the peak due to
deep penetration of light reaction products like protons or alphas. These trends have been 
already seen in MCHIT calculations for light and medium-weight nuclei 
in water~\cite{Pshenichnov:etal:2008}. The non-trivial shape of the depth-dose curves 
shown in Fig.~\ref{fig:Fe_969_1087_Water} is specific for deep-penetrating high-energy 
nuclear beams.   

Moreover, the ionization energy loss, proportional 
to the square of the nuclear charge drops dramatically due to  
nuclear fragmentation since $(Z_1+Z_2+...+Z_N)^2 > Z_1^2+Z_2^2+...Z_N^2$. 
Nuclear evaporation usually leaves a heavy residue, while nuclear 
multifragmentation creates several fragments of
comparable mass. In the latter case the total ionization energy loss of secondary fragments is
reduced considerably compared to typical energy loss of evaporation residues.

These considerations explain the trends seen in Fig.~\ref{fig:Fe_969_1087_Water},
in particular, why the linear energy deposition calculated with accounting for multifragmentation 
is  reduced compared to the calculations, where light particle evaporation is the only 
de-excitation process. However, as seen in Fig.~\ref{fig:Fe_969_1087_Water}, neglecting
multifragment decays gives only very small changes in the depth-dose distributions calculated
both with the cascade and abrasion models. This reflects the fact that the contribution of 
multifragmentation  to the total reaction cross section is small.

It should be stressed, that the calculations with the cascade model overestimate the
linear energy deposition from $^{56}$Fe nuclei in water, while the calculations with the
abrasion model agree very well with the experimental data at two beam energies.
We attribute this deficiency to the fact
that the Light-Ion Binary cascade model underestimates either the fragmentation 
probability of heavy-nuclei, or the excitation energy of nuclear spectators 
produced after the fast stage of the reaction.

However, as also seen from Fig.~\ref{fig:Fe_969_1087_Water}, the results obtained with 
the abrasion model for 969.8 $A$ MeV Fe  underestimate the dose 
after the distal edge of the Bragg peak. This analysis demonstrates the necessity to improve 
physical models for the description 
of the initial stage of high-energy nucleus-nucleus collisions especially for medium-weight 
and heavy projectiles.

\subsection{Yields of secondary fragments produced by medium-weight nuclei in extended media}

\begin{figure}[htbp]  
\begin{centering}
\includegraphics[width=0.85\columnwidth]{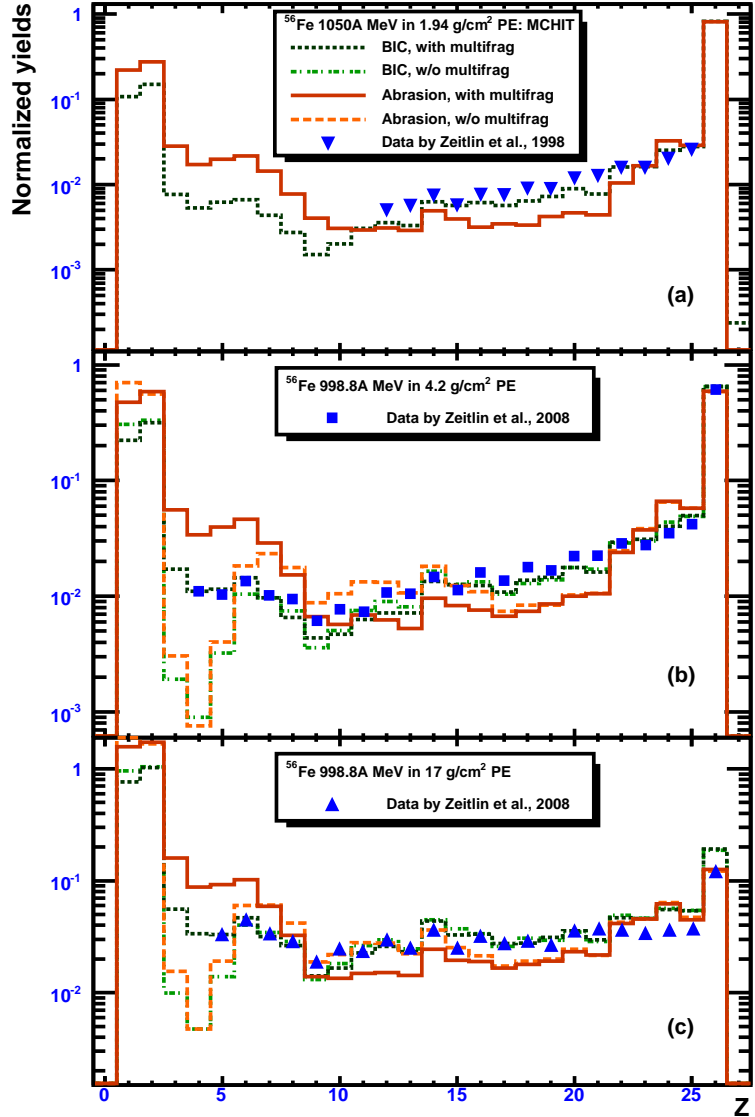}
\caption{Color online. Yields of charged nuclear fragments produced by 1050 $A$ MeV and 998.8 $A$ MeV 
$^{56}$Fe nuclei in polyethylene targets of (a) 1.94, (b) 4.2 and (c) 17 g/cm$^2$ areal densities.
The MCHIT results obtained with the Light-Ion Binary cascade  
model of Geant4 with and without multifragmentation are shown by the dotted and dash-dotted 
histograms, with the Wilson abrasion model with and without multifragmentation 
are shown by the solid and dashed
histograms, respectively. The experimental data from 
Refs.~\cite{Zeitlin:Sihver:et:al:2008,Zeitlin:Heilbronn:Miller:1998} are shown by various points.}
\label{fig:Fe_1GeV_thin_thick_PE_BIC_Abrasion}
\end{centering}
\end{figure}

Further validation of the nucleus-nucleus collision models of the Geant4 toolkit can be performed
with the data on the yields of secondary fragments produced by nuclear beams in thin and thick 
targets. In particular, such data were presented in 
Refs.~\cite{Zeitlin:Sihver:et:al:2008,Zeitlin:Heilbronn:Miller:1998} for $^{56}$Fe nuclei
transversing polyethylene (PE) targets. Calculations with the MCHIT model were performed for
such polyethylene targets with areal densities of 1.94, 4.2 and 17 g/cm$^2$. 
The yields of charged nuclear fragments 
leaving such targets at any angle were scored according to their charge. 
The results were divided by the number of primary beam nuclei to obtain normalized fragment yields  
which were plotted in Fig.~\ref{fig:Fe_1GeV_thin_thick_PE_BIC_Abrasion} together with the
data of Refs.~\cite{Zeitlin:Sihver:et:al:2008,Zeitlin:Heilbronn:Miller:1998}. Again, the cascade and 
abrasion models were used in the MCHIT calculations. 

In Fig.~\ref{fig:Fe_1GeV_thin_thick_PE_BIC_Abrasion} calculations with and without 
multifragment decays are presented for  4.2 and 17 g/cm${^2}$ polyethylene targets making possible to 
estimate the impact of explosive fragmentation on fragment yields predicted by the two 
nucleus-nucleus collision models. As seen from
Figs.~\ref{fig:Fe_1GeV_thin_thick_PE_BIC_Abrasion}(b)  and  
\ref{fig:Fe_1GeV_thin_thick_PE_BIC_Abrasion}(c), the yields of intermediate mass fragments 
with $Z=6-9$ can not be properly described by either of two models
without taking into account multifragment decays.
 
The general shape of $Z$-distributions is better reproduced by the cascade model in combination 
with G4StatMF. The abrasion model with or without multifragmentation underestimates
the fragment yields in a broad range of charges from 10 to 22.      
Since the largest fragments contribute the main part to the total dose, the models should especially
accurately estimate the production of fragments close in mass to the beam nuclei. As seen from 
Fig.~\ref{fig:Fe_1GeV_thin_thick_PE_BIC_Abrasion}(b), both models describe
the yields of $Z=25$ and $Z=26$ nuclei on 4.2 g/cm${^2}$ target. 
However, the cascade model significantly 
overestimates such yields on a more thick target of 17 g/cm$^{2}$, see  
Fig.~\ref{fig:Fe_1GeV_thin_thick_PE_BIC_Abrasion}(c), while the abrasion model is closer
to experimental data for this case. This may explain why the 
abrasion model better describes the depth-dose distributions of Fe nuclei in 
thick targets made of light materials,
as it was already demonstrated in Sec.~\ref{FeDepthDose} for water.

\section{Conclusions}\label{Conclusions}

We have performed validation checks of the C++ versions of the Fermi break-up model 
(G4FermiBreakUp) and the Statistical Multifragmentation model (G4StatMF)  
of the Geant4 toolkit (version 9.1) by comparing their results for
multifragmentation of light, medium-weight and heavy nuclei with results of the FORTRAN
versions of the corresponding models and with 
available experimental data. Two versions of the Fermi break-up model agree well
after the corrections to the energies of excited levels 
of light nuclei have been made in G4FermiBreakUp.    

After several fixes and updates have been introduced into G4StatMF, we have also achieved 
a rather good agreement with the results obtained with the FORTRAN SMM code. 
The validity of G4StatMF is confirmed in the stand-alone tests,
which simulated the decay of hot nuclear systems with masses between Sn and Pb.
We have also shown that the G4StatMF is able to describe central $^{129}$Xe+$^{\rm nat}$Sn collisions at
32 $A$ MeV. The results of these tests and the proposed corrections were made available to the members 
of the Geant4 collaboration.  The patches to G4StatMF were  officially accepted and became a part of the
Geant4 version 9.2 in December 2008. Now G4StatMF can be used to predict the production of 
various fragments and their isotope distributions in decay of hot medium-weight and heavy nuclei
created in nuclear fragmentation reactions in extended media. 

As shown in our previous 
publications~\cite{Pshenichnov:etal:2005,Pshenichnov:etal:2006,Pshenichnov:etal:2007,MCHIT},
the Light-Ion Binary cascade model is rather successful in 
describing the transport of light nuclei of therapeutic energies in water.
The use of the Wilson abrasion model for such calculations is also validated by the present study.
These models combined with the Fermi break-up model of Geant4 describe well the total dose and
the yields of secondary fragments (H, Li, Be) from carbon beams. 
At the same time future work is needed to improve the description of He and B fragments by these
models. We have to stress that not only theoretical models but also the measurements in 
extended targets should be improved in the future, as currently available experimental data still have
large uncertainties. 

The propagation of energetic Fe nuclei in water has been simulated using different models from 
the Geant4 toolkit. The initial stage of nucleus-nucleus collisions was described either 
by means of the Light-Ion Binary cascade model or by the Wilson abrasion model. 
As pointed out by developers of the cascade model, 
at least one of the colliding nuclei should be lighter than nitrogen. 
This condition is violated for Fe nuclei in water (H$_2$O) that may be the reason of 
overestimated dose before the Bragg peak, possibly due to inaccurate estimation of sizes and  
excitation energies of Fe spectators by this model. There are no restrictions on the masses
of colliding nuclei in the Wilson abrasion model, and this model seems more relevant to
Fe fragmentation in water providing much better agreement with the depth-dose data before the Bragg peak.
However, the dose behind the Bragg peak is underestimated by the abrasion model, possibly, due to
underprediction of the yields of secondary protons and light nuclei.      

Our analysis shows that the Light-Ion Binary cascade model is not good enough 
to simulate fragmentation of heavy nuclei, despite the fact that its component, 
G4StatMF, is valid for stand-alone 
multifragmentation calculations. We conclude that better models for the initial stage
of nucleus-nucleus collisions are needed to describe the transport of heavy nuclei 
in extended media by means of the Geant4 toolkit.  
As shown in Ref.~\cite{Kameoka:et:al:2008}, JQMD (JAERI
version of Quantum Molecular Dynamics) model can be considered as a promising candidate. 
We believe that G4FermiBreakUp and G4StatMF, 
which have been validated in the present paper, can be coupled with such a model
in the future. 

This work was partly supported by Siemens Medical Solutions.
We are grateful to Hermann Requardt for stimulating discussions.
This work is a part of IAEA's Co-ordinated Research Project ``Heavy charged-particle
interaction data for radiotherapy''. The discussions with Diether Schardt, Jos\'e Manuel Quesada and  
Niels Bassler are gratefully acknowledged. 
The authors are grateful to  Michael F. Moyers for providing us the tables of experimental data 
for the depth-dose distributions of Fe nuclei in water.

\end{document}